# Photoinduced insulator-metal transition and nonlinear optical response of correlated electrons — a DMFT analysis

**N. Tsuji, T. Oka, H. Aoki**

Department of Physics, University of Tokyo, Hongo, Bunkyo-ku, Tokyo 113-0033, Japan

E-mail: tsuji@cms.phys.s.u-tokyo.ac.jp

**Abstract**. We investigate a photoinduced insulator-metal transition in the Falicov-Kimball model with the dynamical mean-field theory (DMFT) extended to nonequilibrium systems in periodic modulations in time. When the photon energy of the pump light is $\Omega \approx U$ ($U$: the interaction strength), a Drude-like peak is found to grow in the optical conductivity spectrum, which is an evidence that the system is driven into a metallic state. During the transition, the band gap does not collapse, whereas the distribution function exhibits a non-monotonic behaviour away from the Fermi distribution. This indicates that the transition cannot be accounted for by heating effects, but creation of photo-carriers is responsible.

## 1. Introduction

A photoinduced insulator-metal transition (IMT) opens up a new frontier for controlling physical properties of correlated electron systems, as recently observed in pump-probe spectroscopy experiments [1]. In the transition, carriers are created in the system by irradiation of an intense pump laser (with the frequency $\Omega$), which acts as a driving force of the transition. One big issue is whether the phenomenon can be ascribed to a heating effect picture, *i.e.* an increased effective temperature $T_{eff}$ and the associated Fermi distribution $f(T_{eff})$, or to a conventional band-filling controlled IMT picture [2], where creation of photo-carriers is considered in a similar manner to chemical doping. Whether those pictures are appropriate for the photoinduced IMT is still an open question. To answer the question is the motivation of the present study.

In correlated electron systems we have to deal with two essential ingredients in a theoretical study of the photoinduced IMT: an electron correlation effect and a nonlinear electric-field effect. One theoretical tool that can treat this is the dynamical mean-field theory (DMFT), which maps a lattice model into an effective impurity problem to be solved self-consistently. Recently the method has been extended to nonequilibrium problems [3]. For systems in laser lights, we have to deal with ac fields, for which we have recently proposed [4] to combine the nonequilibrium DMFT with the Floquet method [5] (Floquet + DMFT) to treat nonequilibrium steady states driven by pump lights.

In this paper, we apply Floquet + DMFT to the Falicov-Kimball (FK) model (Sec. 2). The reason for adopting the model is that it can be solved exactly within the method, which will help us build a firm basis for our understanding of nonequilibrium physics. Then in Sec. 3 we present our numerical results for $\Omega \approx U$, a region of interest. Finally we discuss how the system undergoes IMT due to the application of the pump light in Sec. 4, where the $\Omega << U$ case is also shown to be of interest.

## 2. Model and method

Our interest resides in excited states realized during the irradiation of the pump laser. There, even in a steady nonequilibrium state, we asset that energy dissipation, occurring through channels such as phonons or spins, should be an important factor. Here we assume that the system is pumped continuously by the light and is in a steady state due to the balance between the driving field and dissipation. The Hamiltonian of the FK model coupled to the ac pump field is given by

$$H(t) = \sum_{k} \varepsilon_{k - E\sin\Omega t/\Omega} c_{k}^{+} c_{k} + U \sum_{i} c_{i}^{+} c_{i} f_{i}^{+} f_{i},$$

where $E$ is the amplitude of the pump light, $\varepsilon_k$ the band dispersion which we assume to be that of the hypercubic lattice, and $c_i^+(c_i)$ and $f_i^+(f_i)$ represent a creation (annihilation) operator of itinerant and localized electrons, respectively. To describe the relaxation, we introduce a phenomenological damping parameter $\Gamma$ ($1/\Gamma \sim$ the relaxation time) to represent the energy dissipation rate [6]. We also assume that the system relaxes to the thermal equilibrium with temperature $T$.

We have solved the FK model in ac fields exactly with Floquet + DMFT. The details of the theoretical techniques are described in Refs. [4]. The outline of the method is the following: We assume that the driven system is periodic in time in a steady state. Then we can apply the Floquet theorem, which is a time-analog of the Bloch theorem for external fields periodic in space. The Floquet theorem gives some general properties of a solution of the time-dependent Schrödinger equation. Employing the theorem, we can transform Green's functions into a Floquet matrix form [4]. In the same way all the self-consistent equations of DMFT can be expressed in the Floquet form. The advantage of using the Floquet matrix form is that the size of each Floquet matrix that we need to obtain for accurate results can be made quite small, since higher-order processes coming from off-diagonal components of the Floquet matrix of a large size become irrelevant. This fact dramatically reduces the computational time of numerical calculations. With the Floquet + DMFT technique we can treat nonequilibrium steady states beyond the linear-response regime.

## 3. Results

Physical quantities can be calculated from the Green function. Let us denote the $n$-th Floquet mode of the Green function by $(G_k)_n(\omega)$ [4]. The time-averaged density of states (DOS) is given by $A_0(\omega) = -\sum_k \text{Im}(G_k^R)_0(\omega)/\pi$ while the time-averaged density of occupied states by $N_0(\omega) = -i\sum_k (G_k^<)_0(\omega)/2\pi$ (Fig. 1, the left panel). Using these quantities, we define the effective distribution function as $f_{eff}(\omega) = N_0(\omega)/A_0(\omega)$ (Fig. 1, the centre panel). We also calculate the real part of the optical conductivity $\sigma(\omega)$ (Fig. 1, the right panel) through the relation [6]

$$\sigma(\omega) = \frac{1}{\omega}\text{Re}\sum_{k}\frac{1}{\tau}\int_{0}^{\tau}d\bar{t}\int_{0}^{\infty}dt\, e^{i\omega t}\left\langle [j_k(\bar{t}+t/2), j_k(\bar{t}-t/2)]\right\rangle \qquad (1)$$

with $\tau = 2\pi/\Omega$, $[\,,\,]$ the commutator, $\langle\cdots\rangle$ the statistical average, and $j_k(t)$ the current operator.

## 4. Discussion

Let us first examine the behaviour of DOS. When the pump light is absent (Fig. 1 (a)), we observe a clear band gap between the lower and upper bands, where the electrons occupy the lower band only and the system is an insulator. As we increase the amplitude of the pump light, electrons are gradually excited to the upper band (Fig. 1 (b), (c)), which indicates the production of photo-carriers. We can, however, note that DOS is not modified significantly, so that *the band gap does not collapse with photo-irradiation*, a difference between the photoinduced and filling-controlled IMT.

Next, let us consider the effective distribution function. In equilibrium (Fig. 1 (d)), it reduces to the Fermi distribution function $f(T)$. When we switch on the pump light (Fig. 1 (e), (f)), the distribution deviates from $f(T)$, and shows a novel nonequilibrium distribution. One can see that peak and dip

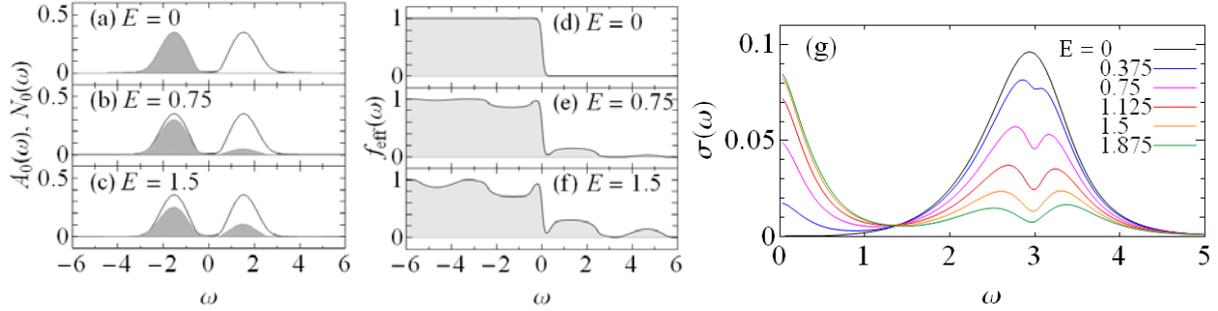

Figure. 1 (a)-(c): The density of states (solid curves) and the density of occupied states (shaded regions). (d)-(f): Corresponding distribution functions. (g): The optical conductivity for $U=3, \Omega=3, \Gamma=0.05,$ and $T=0.05$.

structures appear in the distribution. This is due to photo-excitation with the photon energy $\Omega$, so that the spacing of the structures is nearly equal to $\Omega$. In Fig. 1 (e), (f), it is obvious that the distribution cannot be fitted to an effective Fermi distribution $f(T_{eff})$ with an effective temperature $T_{eff}$ since it is a non-monotonic function. Thus *the heating picture is irrelevant*.

The transition from insulating to metallic states is more clearly seen in the change of the optical conductivity. For small pump electric field $E=0$, it shows an optical gap in the low-energy region, and has a charge transfer (CT) peak around $\omega \sim U$. As $E$ increases, a Drude-like peak structure emerges around $\omega \sim 0$, which is an evidence that the system is driven into a metallic state. One also finds that the pump light reduces the CT peak due to the "bleaching effect". These weight transfers can be explained by the creation of photo-carriers. The fact that the CT peak does not shift towards a lower energy region suggests that the band gap still exists when the system is subject to the pump light.

What is more surprising is that *a new dip structure* appears around $\omega = \Omega$ in the CT peak. We verify [6] that it comes from the contribution of the vertex corrections in Eq. (1). Since the vertex corrections are exactly absent in equilibrium within DMFT, the dip structure originates from quantum correction effects *unique to nonequilibrium*.

Finally, let us consider the case of $\Omega \ll U$. Although usual experiments do not focus on this case, several interesting effects are found. The photon energy is so small compared to the band gap that one-photon absorption/emission processes are forbidden. However, if the intensity of the pump light is sufficiently large, higher-order processes become relevant. For example, the band gap disappears, and the weight of DOS becomes finite in the mid-gap region. Through the photoinduced states in the mid gap region, the electrons are excited from the lower to upper bands, and again photo-carriers are created, which should make the system metallic.

To summarise, we study the FK model driven by the pump light with the Floquet + DMFT method. We capture characteristic properties of photoinduced IMT such as the nonequilibrium distribution, and reveal how it differs from the heating effect or the filling-controlled IMT in equilibrium.